\documentstyle[twoside]{hsproc_nopal}
\newcommand{\SK}{\mbox{Sj\"ostrand-Khoze}}

\newcommand{\npp}{\mbox{$N^{++}$}}
\newcommand{\ndel}{\mbox{$\delta^{++}$}}
\newcommand{\nmc}{\mbox{$M_C^{++}$}}
\newcommand{\mcppp}{M_C^{+++}(Q_3)}

\newcommand{\nppp}{n^{+++}}
\newcommand{\nppn}{n^{++-}}
\newcommand{\nbppp}{N^{+++}(Q_3)}

\newcommand{\delppp}{\delta^{+++}_2(Q_3)}
\newcommand{\delppn}{\delta^{++-}(Q_3)}
\newcommand{\delpppt}{\delta^{+++}(Q_3)}
\newcommand{\delpppg}{\delta^{+++}_{genuine}(Q_3)}

\newcommand{\rthree}{R_3(Q_3)}
\newcommand{\ronetwo}{R_{1,2}(Q_3)}
\newcommand{\cthree}{C_3(Q_3)}

\newcommand {\Zzero}   {{\mathrm Z}^0}

\newcommand{\Gw}  {\Gamma_{\mathrm{W}}}
      
\newcommand{\epem}{\mbox{$\mathrm{e^+e^-}$}}

\newcommand{\WW}{\mbox{$\mathrm{W^+W^-}$}}

\newcommand{\eeWW}{\mbox{\epem$\rightarrow$\WW}}

\newcommand{\qqprime}{\mbox{$\mathrm{q\overline{q}^\prime}$}}

\newcommand{\lnu}{\mbox{$\ell\overline{\nu}_{\ell}$}}
\newcommand{\lpnu}{\mbox{$\ell^+{\nu}_{\ell}$}}
\newcommand{\lmnu}{\mbox{${\ell^\prime}^- \overline{\nu}_{\ell^{\prime}}$}}

\newcommand{\WWqqqq}{\mbox{\WW$\rightarrow$\qqprime\qqprime}}
\newcommand{\WWqqln}{\mbox{\WW$\rightarrow$\qqprime\lnu}}

\newcommand{\WWlnln}{\mbox{\WW$\rightarrow$\lpnu\lmnu}}
\newcommand{\QQQQ}{\mathrm{4q}}
\newcommand{\QQLV}{\mathrm{qq}\ell\nu}

\newcommand{\nchQQQQ}{\langle n_{\mathrm{ch}}^{\QQQQ}\rangle}
\newcommand{\nchQQLV}{\langle n_{\mathrm{ch}}^{\QQLV}\rangle}
\newcommand{\Dnch}{\Delta\langle n_{\mathrm{ch}}\rangle}
\newcommand{\DQQQQ} {\mathrm{D^{\QQQQ}}}
\newcommand{\DQQLV} {\mathrm{D^{\QQLV}}}
\newcommand{\DD} {\Delta\mathrm{D}}
\begin{document}
\titlepage
\title{
\begin{flushright}
  Tel-Aviv HEP Preprint\\ 
  TAUP-2535-98\\
  16  November 1998 \\
~~~~~~~~~~~~~~~~~~~~~~~~~~~~ {}\\  
\end{flushright}}
\begin{center}
%{\Large \bf Bose-Einstein Correlations and Colour Reconnection} \\ 
{\Large \bf Measure of $\pi$'s and $\Lambda$'s emitter radius via}\\
\vspace*{2mm}
%{\Large \bf in W-pairs and Hadronic Events at LEP}
{\Large \bf Bose-Einstein and Fermi-Dirac Statistics}\footnote{
Based on an invited talk given
 by G. Alexander in the 
 Hadron Structure '98, $7-13$ Sept. 1998, Stara Lesna, Slovakia (to be 
published in this Conference Proceedings).}
\end{center}

\vspace*{3mm}

\author{Gideon Alexander and Iuliana Cohen}
{School of Physics and Astronomy, Tel-Aviv University, Tel-Aviv, Israel }

\smallskip

\abstract{ This report summarises the recent studies of the Bose-Einstein
Correlations (BEC) carried out by the four LEP experiments using 
hadronic Z$^0$ decay events and $e^+ e^-$ reactions leading to the
WW decay final states. The three identical charged pion systems 
$\pi^{\pm} \pi^{\pm} \pi^{\pm}$ have been studied by OPAL with 
$4  \times 10^6$ hadronic Z$^0$ decays events.
The genuine three-pion correlations corrected for Coulomb effects, 
were separated from the known two-pion correlations by a 
novel subtraction procedure using the data events themselves.
From this analysis a value of $\mathrm r_3 = 0.580 \pm 0.029 ~fm $ was 
obtained.
The recent L3 study of the two-dimensional two-pion BEC 
is also discussed. In this study, which utilised the longitudinal
co-moving system, evidence is given for a non-spherical 
di-pion emitter shape where the ratio of the minor 
to major axes is found to be  $0.74 \pm 0.09$.
Next we summarise the experimental aspects of the BEC and
colour reconnection phenomena in W-pair events at LEP and their contribution
to the systematic error associated with the M$_W$ measurements. Finally we
describe the first attempts to evaluate the   
$\Lambda\Lambda$($\bar{\Lambda}\bar{\Lambda}$) emitter dimension via
the Pauli exclusion principle with a combined result of  
 $ \mathrm R_{\Lambda\Lambda} = 0.14 ^{+ 0.07}_{-0.03} ~~{\mathrm fm} $.}

\vspace{4mm}

\section{Introduction}

\vspace*{4mm}

\noindent
Bose-Einstein Correlations (BEC) of two identical bosons, in
particular two pions and two kaons, have been studied in the past
over a wide range of energies in various particles interactions.
From these measured correlations it was possible to extract values 
for the dimension of the bosons emitter which in the case of $e^+ e^-$
annihilation was found to be in the range
of 0.7 to 1.0 fm independent of energy. The $e^+ e^-$ collider LEP
at CERN, which operated for several years at the energy of the
Z$^0$ mass, resulted in some 4$\times 10^6$ hadronic events collected in
each of the four detectors, ALEPH, DELPHI, L3 and OPAL. These high statistics 
samples have been recently utilised to extend the BEC studies to higher orders
and dimensions. In its 2nd phase, the LEP energy was increased in steps
to reach in 1998 the value of
189 GeV, well above the $W^+ W^-$ production threshold. This presented
the experiments with the possibility to measure M$_W$ with high precision 
provided that the final state interactions (FSI) effects of colour 
reconnection (CR) and the two-pion BEC 
are well understood. In this note we summarise in Section 2 the most recent
studies of the genuine three charged pions BEC of OPAL and in Section 3 the
analysis of the 2-dimensional two-pion BEC of L3 carried 
out with hadronic Z$^0$
decay events is described. We further summarise in Section 4 the 
results obtained by the four LEP experiments for the 
colour reconnection and $\pi^+ \pi^+$ BEC effects\footnote{
Throughout this report,
whenever we refer to a specific boson or baryon system 
we also mean its corresponding charge conjugate state.}
in the $e^+ e^- \to W^+ W^-$ reaction. Whereas the emitter radius of
two identical bosons, mainly pions, have evaluated in many reactions
and energies, essentially nothing was known until recently on
the emitter dimension of two identical baryons, $R_{BB}$, in multihadronic
final states. The first measurements of $R_{\Lambda \Lambda}$ in
the hadronic Z$^0$ events are described in Section 5. Finally
conclusions are drawn in Section 6. 

\vspace*{2mm}

\noindent
\section{Studies of the one-dimensional 3$\pi$ BEC} 
%in$\Zzero$ decays at LEP}

\vspace*{4mm}

\noindent
The BEC of two identical charged pions can be measured in terms
of the ratio $R_2(Q_2)$, which in this case is identical to the correlation
function $C_2(Q_2)$, namely:
$$R_2(Q_2) \equiv C_2(Q_2) = \frac{\npp (Q_2)}{\nmc (Q_2)} = 1 + 
\frac{\npp(Q_2) - \nmc (Q_2)}{\nmc (Q_2)}
= 1 + \frac{\ndel (Q_2)}{\nmc (Q_2)}$$
Here $\npp (Q_2)$ is the number of $\pi^+ \pi^+$ data combinations
and $\nmc (Q_2)$ is the corresponding Monte Carlo (MC) combinations where
the MC simulates the data  
in all aspects but is void of BEC.
The Lorentz invariant $Q_2$ variable used in this ratio is defined by
$Q^2_2 = -(q_1 - q_2)^2$ where $q_1$ and $q_2$ are the four momenta
of the two identical pions. In the two-pion BEC studies the values
 for the wo-pions emitter radius were found  to be in the range $0.6 - 1.2$ ~fm,
 independent of the interaction type or the 
centre-of-mass energy \cite{wolf}.  

\noindent
Similarly the three-pion BEC can be studied by the ratio
%\begin{center}
%\begin{equation}
$$\rthree \ =  \ \frac{\nbppp}{\mcppp} \ = \
1 + \frac{\nbppp - \mcppp}{\mcppp} \ 
= \ 1 + \frac{\delpppt}{\mcppp}$$ 
%\end{equation}
%\end{center} 
where the variable $Q_3$ is defined by 
$Q^2_3 = -(q_1 - q_2)^2  -(q_1 - q_3)^2 -(q_2 - q_3)^2$. However,
unlike the case of $R_2(Q_2)$, 
$R_3(Q_3)$ does {\it NOT} only measure the so called ''Genuine'' 
three-pion BEC but also the two-pion BEC and therefore $R_3(Q_3) \ne
C_3(Q_3)$. Thus the measurement of the genuine three-pion correlation,
requires a special procedure to subtract from $R_3(Q_3)$ the 
contribution coming from the two-pion BEC.\\
Assuming a spherical shape with an exponential density for
the three-pion emitter one can parametrise the experimental
distribution by the expression

\vspace*{-2mm}

\begin{center}
\begin{equation}
\label{eq_c3qd}
\cthree \ = \ \kappa (1 + 2\lambda_3 e^{-Q^2_3r^2_3})
(1 + \varepsilon Q_3 + \delta Q_3^2)
\end{equation}
\end{center}
where $\lambda_3$, which can vary within the limits $0 \le \lambda_3
\le 1$, measures the strength of the three-boson BEC effect and $r_3$
estimates the size of the three-boson emitter.
The $\kappa$ is a normalisation factor and the term $(1 +
\varepsilon Q_3 + \delta Q_3^2)$ accounts for the long range
correlations arising
from charge and energy conservation and phase space constraints.  
\begin{figure}[htbp]
\begin{minipage}[0.5]{6.5cm}
\begin{center}\mbox{\input epsf \epsfysize 7.0cm
                        \epsfbox{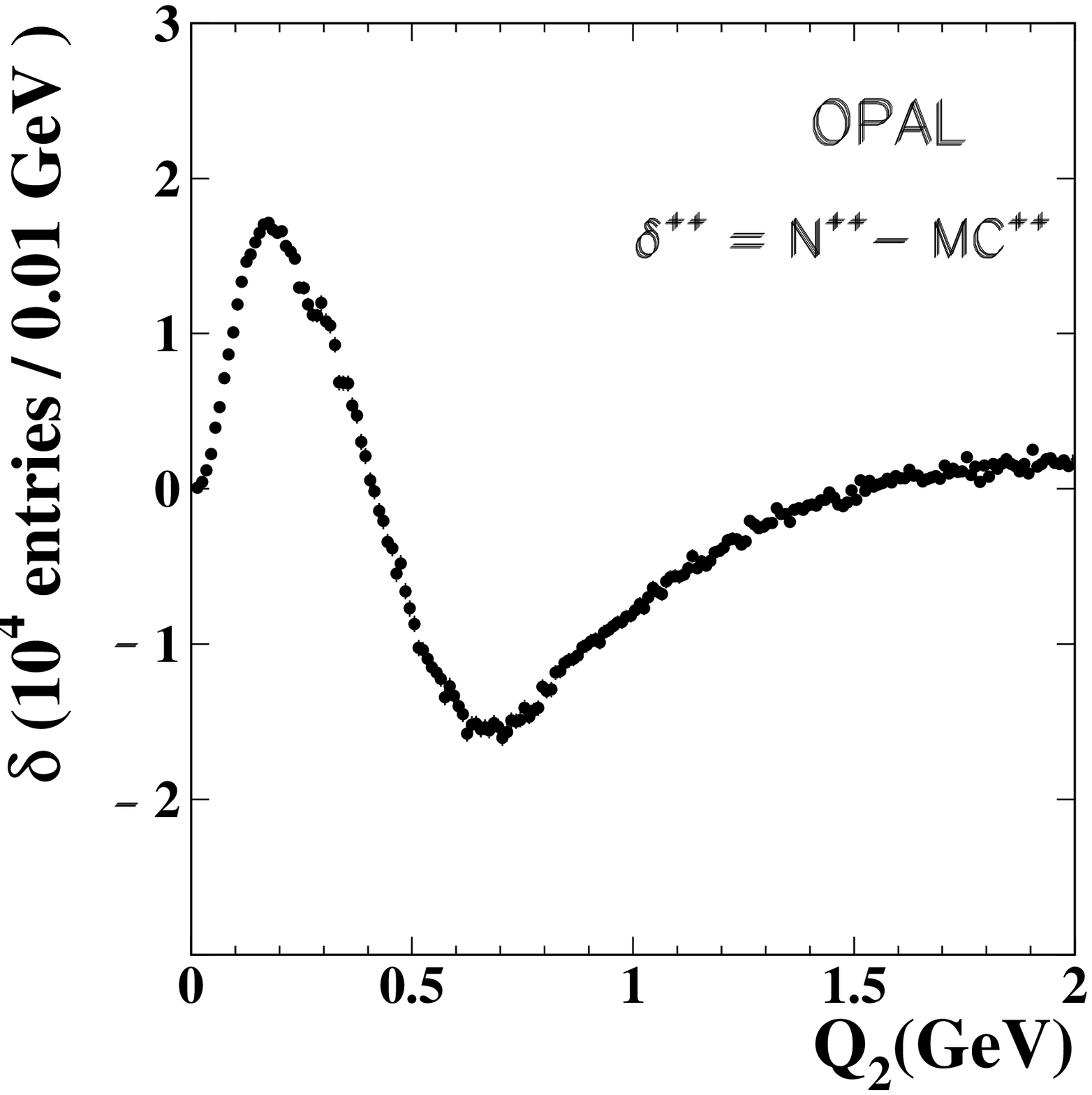}}\end{center}
\end{minipage}
\begin{minipage}[0.5]{6.5cm}
\begin{center}\mbox{\input epsf \epsfysize 7.0cm
                        \epsfbox{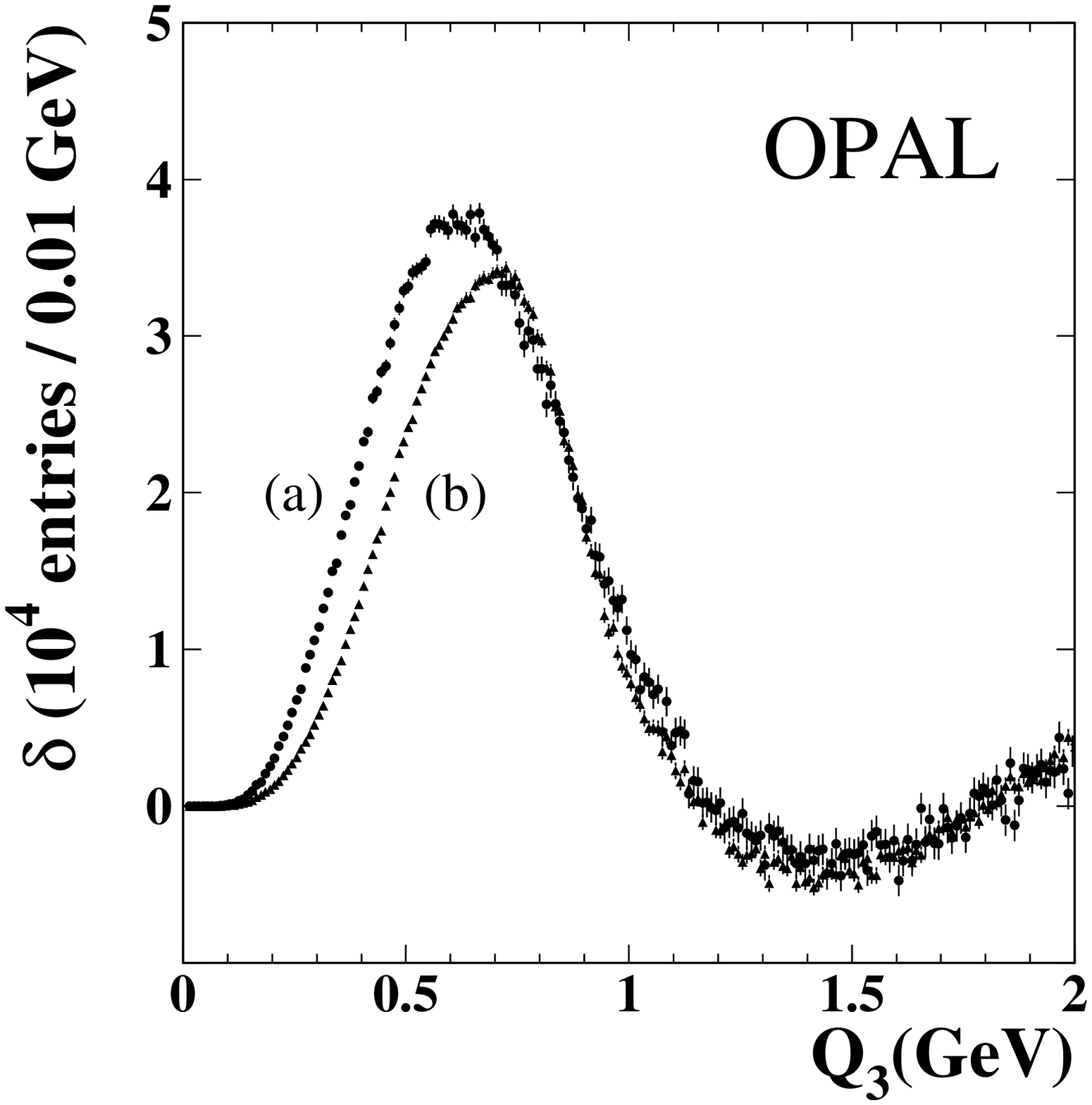}}\end{center}
\end{minipage}
\caption{[left] The excess of two identical pions due to BEC as a 
function of Q$_2$. [right] (a) $\delta^{+++}(Q_3)$ and
(b) $\left < n^{+++}/n^{++-} \right > \times \delta^{++-}(Q_3)$ as a 
function of $Q_3$.}
\label{fig_del}
\end{figure}  

\newpage

\subsection{The genuine 3$\pi$ BEC}
\label{888}

\vspace*{2mm}

\noindent
The genuine Bose-Einstein Correlations (BEC) of three identical
charged pions,
$\pi^+ \pi^+ \pi^+$ corrected for the Coulomb effects have been studied 
in $4 \times 10^6$ hadronic Z$^0$ decays recorded by the OPAL  
detector at LEP\@ \cite{opal3pi}. In this study the chosen reference sample
was generated by the JETSET7.4 Monte Carlo (MC) program. To subtract 
from $R_3(Q_3)$ the contribution of the two-pion BEC a novel
method has been applied which utilises the data events themselves.
The over-all excess  $\delta^{+++}(Q_3)$ can be
written as a sum 

$${ \delpppt \ = \ {\delppp} + {\delpppg }}$$

\noindent
where ~$\delppp$ ~is the three-pion enhancement due to the two-pion
BEC. From straightforward combinatorial considerations 
the relation    
$${ 
\int \delppp dQ_3\ =   \left<\frac{\nppp}{\nppn}\right >\times \int \delppn 
 dQ_3} $$ 
strictly holds when summed over all allowed $Q_3$ values. 
The factor $\left<\nppp/\nppn\right >$ is the average ratio between
 the number of $\pi^+\pi^+\pi^+$ and the $\pi^+\pi^+\pi^-$ combinations 
per event of  the
analysed data sample. It was
further verified experimentally that this relation holds also to a very good
approximation 
in its differential form, namely   
$${ \delppp \ = \left<\frac{\nppp}{\nppn}\right > \times \delppn } $$ 
Thus the genuine three-pion correlation function is given by 
$${ 
\cthree \ = \ \rthree - \ronetwo \ =  \frac{\nbppp}{\mcppp} - 
\frac{\delppn}{\mcppp} \times \left <\frac{\nppp}{\nppn}\right > }$$
where $\ronetwo \ = \ \delppp/\mcppp$. In Fig. 1 [right] the data 
distributions of $\delta^{+++}$ and 
$\left<\nppp/\nppn\right > \times \delta^{++-}$ are shown
as a function of $Q_3$. As can be
seen, the two distributions overlap at values of $Q_3 > 0.7$ GeV. At
smaller $Q_3$ values, a significant difference is seen between
the two distributions which is attributed to the genuine three-pion
BEC. 
This $\delta^{+++}$ representation of the 
three-pion BEC enhancement can of course also be used for the 
two-pion BEC as is seen in Fig. \ref{fig_del} [left]
for the OPAL data \cite{2pi}.
The related $R_3(Q_3)$ and $R_{1,2}(Q_3)$ are shown in Fig. 2a
and the genuine three-pion BEC function $C_3(Q_3)$ is shown in Fig. 2b
where a clear enhancement is observed at the lower end of the $Q_3$
range. A fit of Eq. \ref{eq_c3qd} to the data yields for the
genuine three-pion BEC strength $\lambda_3$ and 
emitter dimension $r_3$ 
the values
$$\lambda_3 = 0.504 \pm 0.010(stat.) \pm 0.041(syst.) \ \rm{and} 
\ r_3 = 0.580 \pm 0.004(stat.) \pm 0.029(syst.) \ \rm{fm}$$ 
where the uncertainties of the measured parameters are
strongly dominated by the systematic errors. These values are
also illustrated in Fig. 4 by the 68$\%$ and 95$\%$ confidence
level contours and the best value point drawn in the $r_3 - \lambda_3$
plane.
The shape of the contours is determined
mostly from the systematic errors. 
These results of OPAL establish the existence of a genuine three-pion BEC
with a significance of about 10 s.d.\\ 
It is of interest to compare
the OPAL results with those obtained in a former study of the 
genuine three-pion BEC carried out
by DELPHI \cite{del3pi} with 10$^6$ Z$^0$ decay events.  
In that analysis no account has been taken
for the Coulomb effect and the subtraction of the two-pion BEC has been
achieved with the help of a MC generated sample and an event mixing
procedure. The DELPHI $R_3(Q_3)$ of the $\pi^+\pi^+\pi^+$ and
the $\pi^+\pi^+\pi^-$ states are shown in Fig. 3. The fitted values for the
genuine BEC strength and the emitter dimension are given in Table 1 where
they can be compared with the OPAL results before the inclusion of
the Coulomb correction. Whereas the two experiments agree within 2 s.d.
on the $r_3$ value they differ in their $\lambda_3$ values which in
the case of DELPHI is less than 4 s.d. away from zero. 
The OPAL and DELPHI $r_3$ results are compared in Table 2 with former 
values obtained in three-pion BEC studies in $e^+ e^-$ annihilation
and hadronic interactions. Most of these analyses have not isolated the
genuine three-pion BEC nor did they correct for Coulomb
effects. Nevertheless in general the average $r_3$ value is seen 
to be smaller
than $r_2$ in agreement with the expectation \cite{juricic} that  
$r_2/\sqrt{3} \leq r_3 \leq r_2/\sqrt{2}$ which in the case of
$r_3^{genuine}$ reduces to $r_3^{genuine} = r_2/\sqrt{2}$. 

\vspace*{5mm}
 
\begin{table}[!hbp]
\begin{center}
\begin{tabular}{|l||c|c|}
\hline
Parameter & Without Coulomb Corr. & With Coulomb Corr.
\cr
\hline \hline
{\bf OPAL} & & \cr
$\lambda_3$ & $0.462 \pm 0.012 \pm 0.041$ & $0.504 \pm 0.010 \pm 0.041$
\cr
$r_3$~(fm) & $0.616 \pm 0.005 \pm 0.029$ & $0.580 \pm 0.004 \pm 0.029$
\cr
$\chi^2/d.o.f.$ & 218/171 & 190/171 \cr
\hline \hline
{\bf DELPHI} & & \cr
$\lambda_3$ & $0.28 \ \pm 0.05 \ \pm 0.07$ & 
\cr
$r_3$~(fm) & $0.657 \ \pm 0.039 \ \pm 0.032 $ & 
\cr
\hline
\end{tabular}
\end{center}
\caption{Results from the OPAL and DELPHI genuine three-pion BEC analyses.
The OPAL values of the $\chi^2$ over degrees of freedom (d.o.f) are
also given.}
\label{tab_OD}
\end{table}

\vspace*{9mm}

\begin{figure}[hbtp]
\begin{minipage}[0.5]{6.6cm}
\begin{center}\mbox{\input epsf \epsfysize 6.5cm
                        \epsfbox{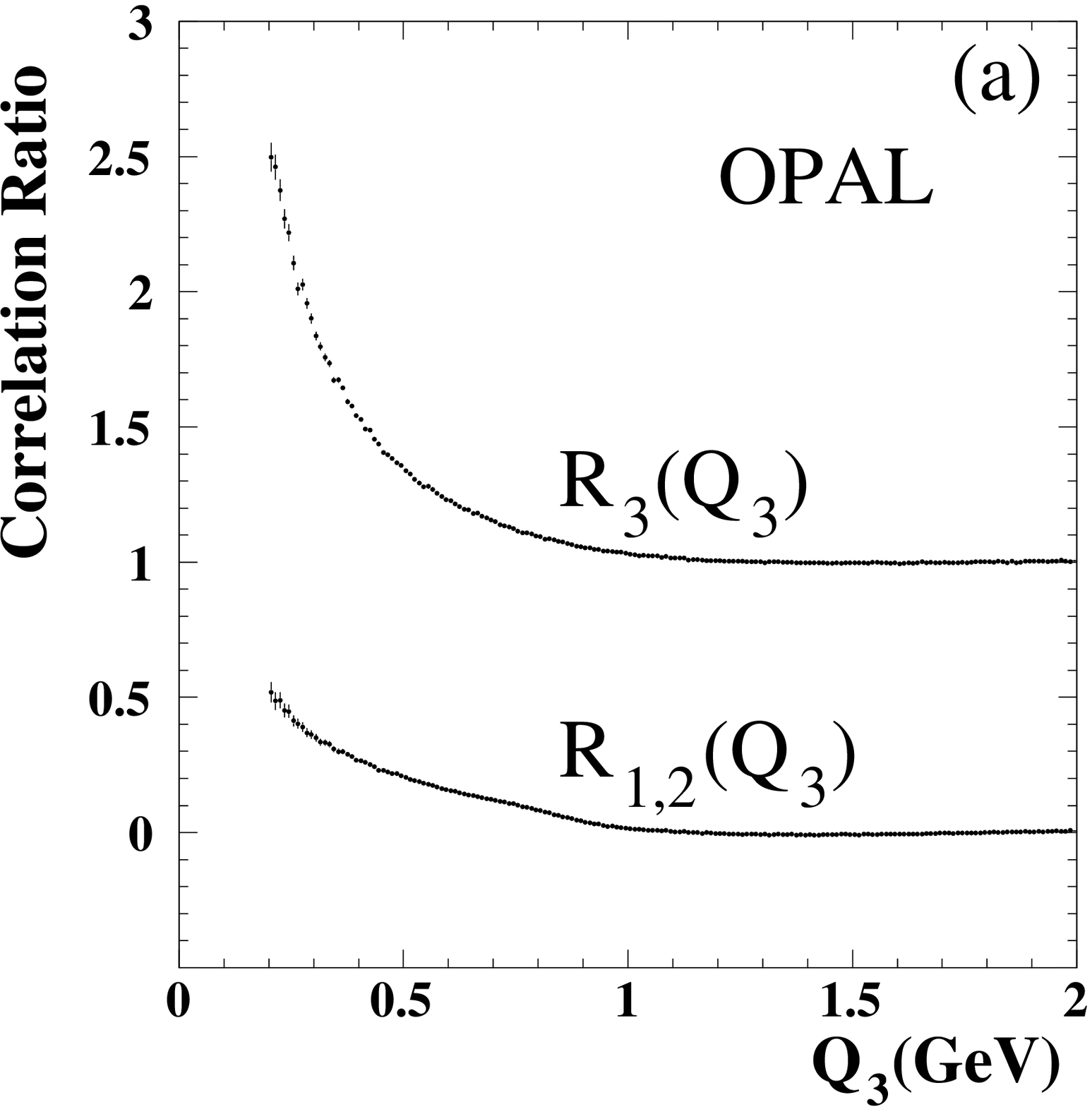}}\end{center}
\end{minipage}
\begin{minipage}[0.5]{6.6cm}
\begin{center}\mbox{\input epsf \epsfysize 6.5cm
                        \epsfbox{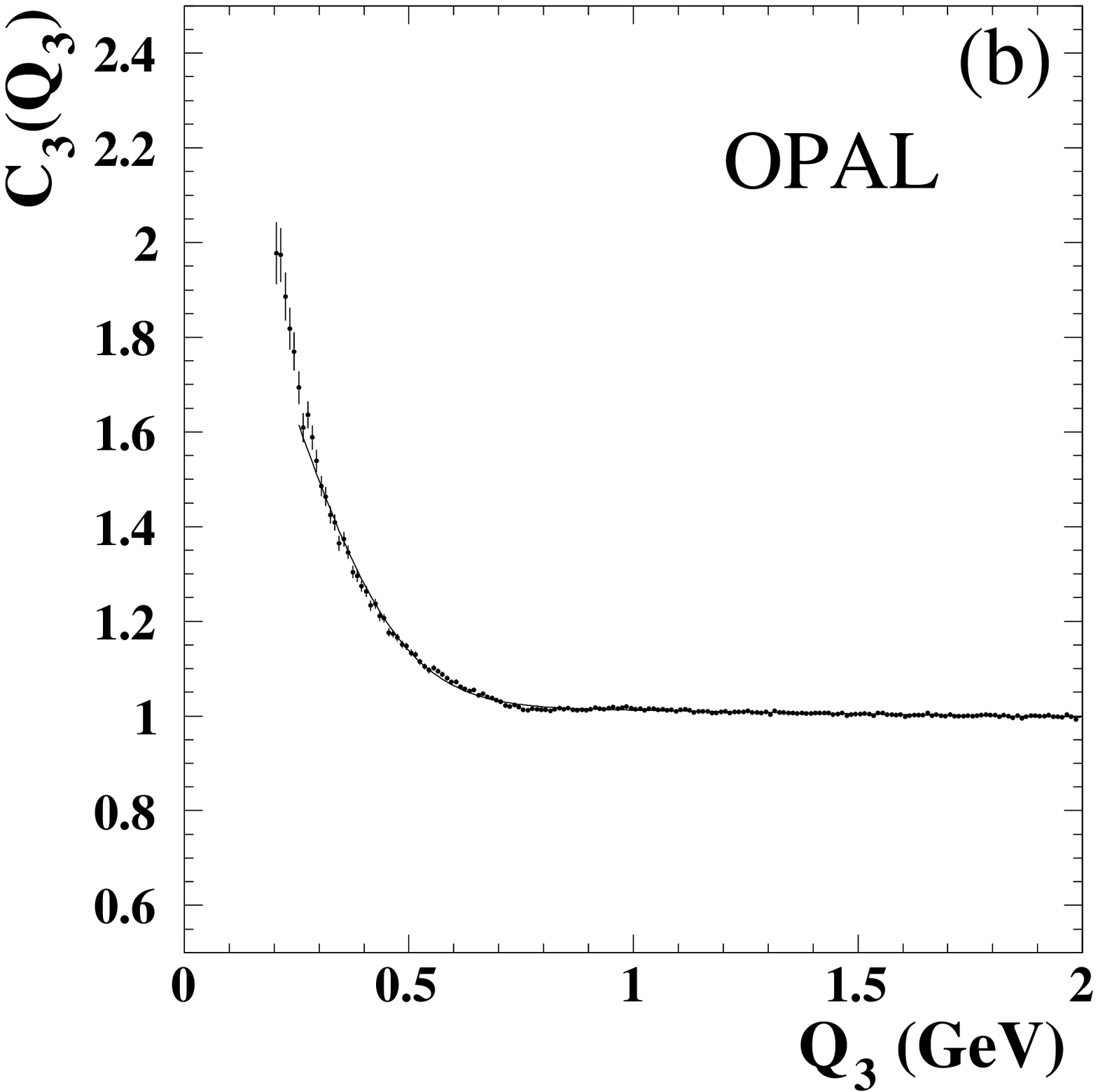}}\end{center}
\end{minipage}
\caption{The Coulomb corrected three-pion BEC distributions measured by OPAL.}
\label{fig_othreepi}
\end{figure}
  
\hspace*{20cm} s

\vspace*{-18mm}

\begin{figure}[hbtp]
\begin{center}\mbox{\input epsf \epsfysize 10.5cm
                        \epsfbox{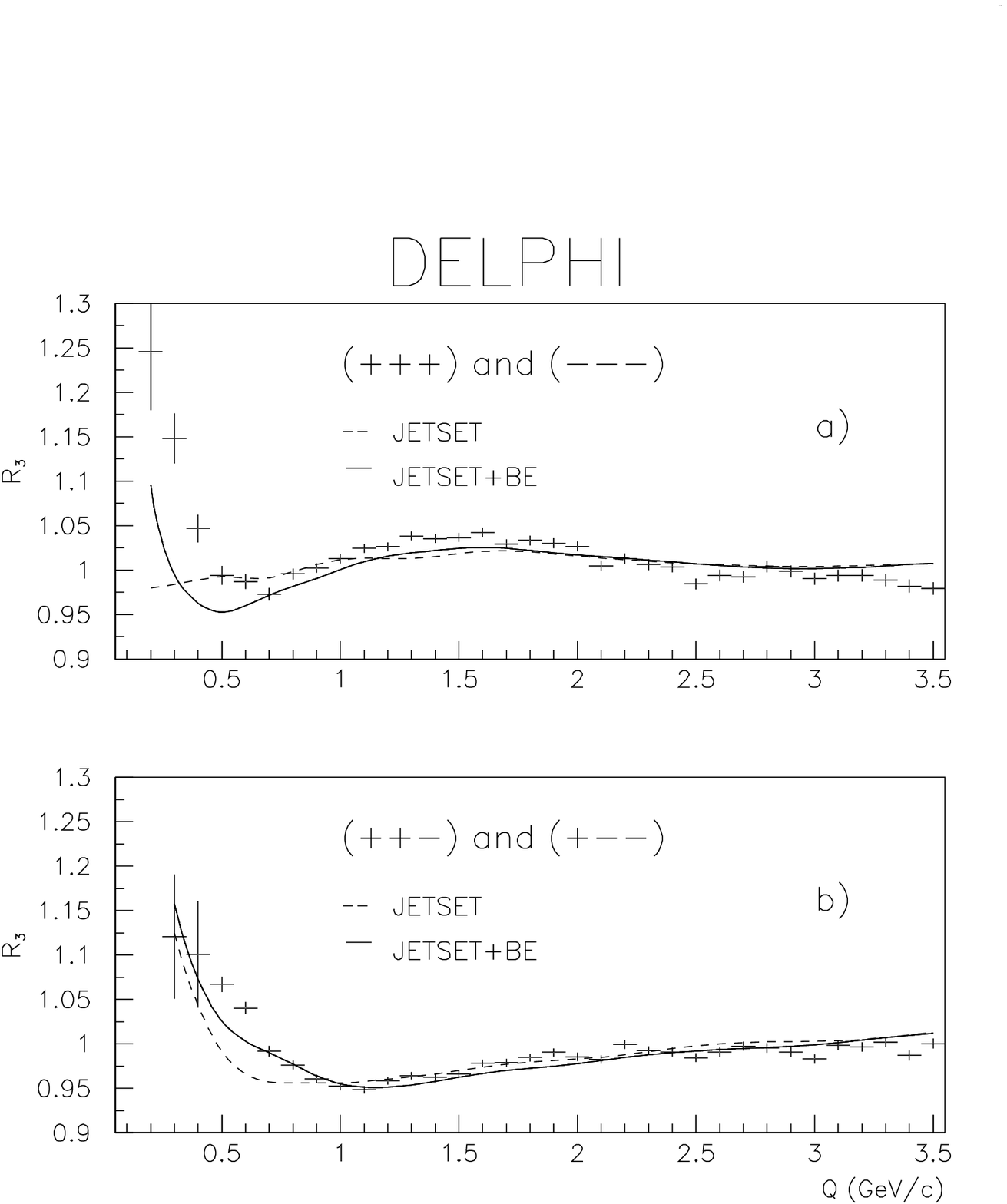}}\end{center}
\caption{The DELPHI measured R$_3(Q_3)$ function for (a) like-sign pion triplets and
(b) unlike-sign pion triplets compared with JETSET7.3 Monte Carlo expectations.}
\label{fig_dthreepi}
\end{figure}

\begin{table}[hbtp]
  \begin{center}
    \begin{tabular}{|l|c|c|c|c|}
      \hline
       Experiment& Reaction & Genuine & Coulomb & ${\bf r_3~(\mbox{fm})}$ 
      \cr
      \hline
%      Heavy Ions (Bevalac) \cite{liu} & Ar Pb & yes & yes &
%      $5.80 \pm 0.75$ & (stat.)
%      \cr
%      Heavy Ions (Bevalac) \cite{liu} & Ar KCL & yes & yes &
%      $4.06 \pm 0.49$ & (stat.)
      AFS  & $p p$ &  {\bf no} & { \bf no} &
      $0.41 \pm 0.02$ ~~(stat.)
      \cr
      NA23 & $p p$ &{ \bf no}  &{ \bf no}  &
      $0.58 \pm 0.07$ ~~(stat.)
      \cr
      NA22  & $\pi^+/K^+ p$ & { \bf yes} & { \bf yes} &
      $0.51 \pm 0.01$ ~~(stat.)
      \cr
      \hline
      TASSO  & $e^+ e^- \to ~q \bar{q}$ & { \bf no} & { \bf no} &
      $0.52 \pm 0.07$ ~~(stat.)
      \cr
      MARK-II (SPEAR)  & $e^+ e^- \to J/\psi$ & { \bf no} & {\bf yes} &
      $0.53 \pm 0.03$ ~~(comb.)
      \cr
      MARK-II (PEP)  & $e^+ e^- \to ~\gamma \gamma$ &{ \bf no} & { \bf yes} &
      $0.55 \pm 0.03$ ~~(comb.)
      \cr
      MARK-II (SPEAR)  & $e^+ e^- \to ~q \bar{q}$ &{ \bf no}  & { \bf yes} &
      $0.45 \pm 0.04$ ~~(comb.)
      \cr
      MARK-II (PEP)  & $e^+ e^- \to ~q \bar{q}$ & { \bf no} &{\bf yes}  &
      $0.64 \pm 0.06$ ~~(comb.)
      \cr
      \hline \hline
      {\bf DELPHI}  & ${ e^+ e^- \to ~{\mathrm Z}^0}$ & {\bf yes} &{ \bf no} &
      ${ \bf 0.66 \pm 0.05}$ ~~(comb.)
      \cr
      {\bf OPAL } & ${\bf e^+ e^- \to ~{\mathrm Z}^0 }$ &{\bf yes} &{\bf yes} &
      ${ \bf 0.58 \pm 0.05}$ ~~(comb.)
      \cr
      \hline
    \end{tabular}
  \end{center}
\caption{Compilation of the three-pion BEC results for r$_3$. The quoted 
statistical errors are marked by (stat.) and  
where the statistical and 
systematic error are combined in quadrature the result in marked by (comb.).}
\label{tab_comp}
\end{table}

\begin{figure}[hbtp]
\begin{center}\mbox{\input epsf \epsfysize 8.5cm
                        \epsfbox{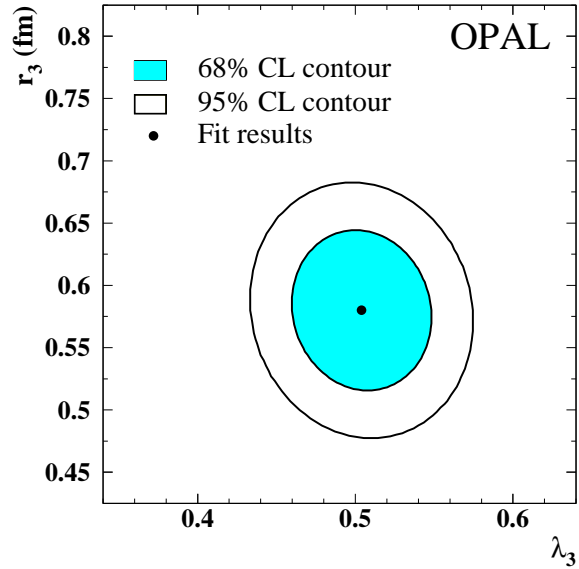}}\end{center}
\caption{The $68\%$ and $95\%$ confidence level contours obtained
by OPAL for the genuine three-pion BEC parameters $\lambda_3$ and $r_3$
after Coulomb correction.} 
\label{fig_contur}
\end{figure}  

\newpage

\section{Two-dimensional 2$\pi$ BEC} 

\vspace*{4mm}

\begin{center}
\begin{figure}[htbp]
\begin{minipage}[0.1]{7.0cm}
\begin{center}\mbox{\input epsf \epsfysize 7cm
                        \epsfbox{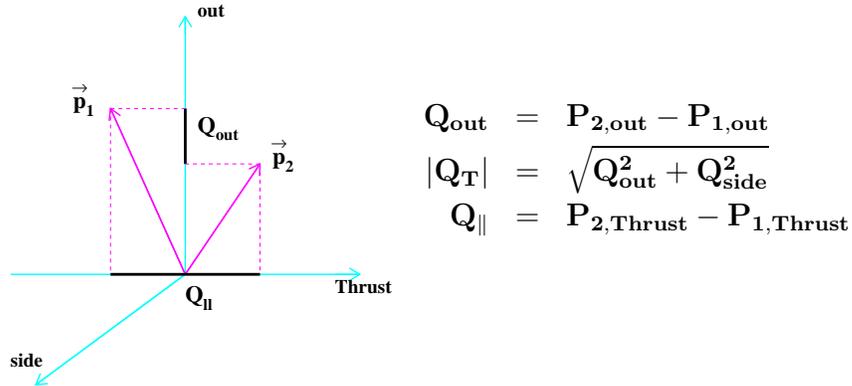}}\end{center}
\end{minipage}
\begin{minipage}[0.1]{6.0cm}
{\large
\begin{eqnarray*}
{\mathbf Q_{out}}& = & {\mathbf P_{2,out} - P_{1,out}}  \\   
{\mathbf |Q_T|}  & = & {\mathbf \sqrt{Q^2_{out} + Q^2_{side}}}  \\   
{\mathbf Q_{\|}} & = & {\mathbf  P_{2,Thrust} - P_{1,Thrust}} 
\end{eqnarray*}
}
\end{minipage}
\caption{LCMS  system drawn for the case where the ${\vec{\mathbf p}}_1$ 
 and $\vec{\mathbf p}_2$ are
in the plane defined by the `out' and `thrust' axes.}
\label{fig_lcms}
\end{figure}
\end{center}

\noindent
In general BEC analyses of the two and more pion systems, like those
discussed in the former subsection, are
parametrised in a one-dimensional emitter shape i.e., 
they are taken to be of a spherical shape. It is however not
unreasonable to expect that in the presence of jet fragmentation
the pion emitter shape will deviate significantly from a sphere.
This possibility has been recently the subject of many 
theoretical studies \cite{2dth}. On the experimental side, 
the L3 collaboration 
\cite{2d} undertook a two-dimensional BEC analysis of the $\pi^+ \pi^+$
pairs using about 10$^6$ hadronic Z$^0$ decay events. 
The analysis used the longitudinal co-moving 
system (LCMS) \cite{lcms}. This system, shown in Fig. \ref{fig_lcms}, 
is defined for each pair of
pions as the system in which the sum of the pion-pair momenta
$\vec{p}_1 + \vec{p}_2$, referred to as the 
`out' axis, is perpendicular to `thrust' axis. The third
axis, which is perpendicular to the `out'$-$`thrust' plane, is referred
to as the `side' axis. For each pion-pair $Q_{out}, Q_{side}$ and 
$Q_{thrust} = Q_{\|}$ are determined. In a two-dimensional analysis
one adds in quadrature  $Q_{out}$ and $Q_{side}$ to form the
transverse component $Q_T$ (see Fig. \ref{fig_lcms}). The Lorentz invariant
variable used in one-dimensional analysis is then equal to
$$Q^2 = Q^2_{\|} + Q^2_T = Q^2_{\|}+Q^2_{out}+Q^2_{side}$$

\vspace*{-8mm}

\noindent
For the BEC analysis L3 has used a reference sample which was created by
the mixing events technique. In the analysis, the experimental 
distributions in the
two-dimensional space $Q_{\|}$ and $|Q_T|$, are parametrised by
\vspace{-4mm}

\begin{center}
\begin{equation}
\label{eq_td}
R_2(Q_{\|}, |Q_T|) = \kappa(1+\delta Q_{\|}+\epsilon |Q_T|)
\left [1 + \lambda e^{ (-R^2_{\|}
   Q^2_{\|} - R^2_T Q^2_T )}\right ] 
\end{equation}
\end{center}

\noindent
where $\lambda$ measures the strength of the BEC effect and $R_{\|}$
and $R_T$ are respectively the parallel and perpendicular dimensions
of the pion emitter. The term
$(1+\delta Q_{\|}+\epsilon |Q_T|)$ takes in account possible long range
correlations and $\kappa$ is a normalisation factor. 
The values of $\lambda$, $R_{\|}$ and  $R_T$ obtained in the
fit of Eq. \ref{eq_td} to the entire data are listed
in Table \ref{tab_2D}. As expected, $R_{\|}$ is larger than 
$R_T$ indicating a non-spherical shape for the pion emitter.
It has been suggested  that the radii $R_{\|}$ and $R_T$  
are function of the two-pion transverse mass
$m_{\mathrm T} = \frac{1}{2}[(m^2+p^2_{t1})^{1/2} + (m^2+p^2_{t2})^{1/2}]$
namely, the higher $m_T$ the lower are the radii \cite{2dth}.
Furthermore in heavy ion collisions the relation between
the radii and $m_T$ was found to be $R \propto 1/\sqrt{m_T}$.
As for $\lambda$, it is expected to increase with $m_T$.    
For this reason L3 repeated their analysis fitting Eq. \ref{eq_td} to
the data in three separated $m_T$ regions. The results of these
fits are shown in Table  \ref{tab_2D}. Finally the fitting procedure
has also been repeated for the subsample of 2-jet hadronic Z$^0$ decay
events (not shown). The results of this two-dimensional BEC analysis
show that:\\
a) The pion emitter is not spherical with radii ratio of 
$R_T/R_{\|} = 0.74 \pm 0.09$;\\
b) General decrease of $R_{\|}$ and $R_T$ is observed as $m_T$ increases;\\
c) The increase of $\lambda$ with $m_T$ cannot be confirmed;\\
d) No significant difference is observed between the 2-jet and the
total hadronic Z$^0$ decay samples.\\
\begin{table}[htbp]
\begin{center}
\begin{tabular}{|c|c|c|c|}
\hline
Range~(GeV) & $\lambda$ & { R}$_\|$~(fm)  & { R}$_\bot$~(fm) \\
\hline \hline
{total sample}         & $0.30 \pm 0.01 \pm 0.07$ & $0.73 \pm 0.02 \pm 0.04$ & $0.54 \pm 0.02 \pm 0.05$  \cr
\hline
$0.14 < m_T < 0.30$ & $0.30 \pm 0.02 \pm 0.06$ & $0.76 \pm 0.05 \pm 0.08$ & $0.72 \pm 0.04 \pm 0.04$  
\\
$0.30 < m_T < 0.45$ & $0.40 \pm 0.02 \pm 0.06$ & $0.70 \pm 0.04 \pm 0.02$ & $0.51 \pm 0.03 \pm 0.03$
\\
$ m_T > 0.45 $       & $0.35 \pm 0.03 \pm 0.06$ & $0.63 \pm 0.04 \pm 0.04$ & $0.38 \pm 0.02 \pm 0.02$
 \\
\hline \hline
\end{tabular}
\end{center}
\caption{Results from the L3 two-dimensional BEC analysis.}
\label{tab_2D}
\end{table}

%\newpage                         
\section{FSI effects on the M$_W$ measurement in \eeWW}

\vspace*{4mm}

\noindent
In the first two years of the LEP2 operation about 75 pb$^{-1}$
of data have been recorded in each of the four LEP experiments.
Out of these 10 pb$^{-1}$ were recorded at $\sqrt{s}$ = 161 GeV,
just above the $W^+ W^-$ threshold, 
where the mass of the W was evaluated from the production cross section. 
The rest of 
the data, taken at 172 and 183 GeV, were used for the M$_W$ measurement
through a direct W reconstruction from the hadronic final states.\\
The \eeWW ~reaction can be divided into three samples 
characterised by their decay channels. These are given in Table \ref{tb_ww} 
together with the corresponding
branching ratio (BR) values and the average values for their purity
and detection efficiency in the LEP experiments. The first item 
consists of a pure 
hadronic event sample where FSI can occur within a single W decay 
or between the two different W bosons. In the second sample, 
often referred to as
the semileptonic event sample, final state interaction can only occur
within a single W and thus is not expected to affect the direct W mass
reconstruction. In the full leptonic events sample no
direct M$_W$ reconstruction is possible. \\
From the practical point
of view of the M$_W$ measurement, FSI are only of consequence if
they occur between the two W's in the hadronic events. 
A common used method for the W mass measurement with the
selected $e^+ e^- \to \WWqqqq$ events consists of grouping the final state 
hadrons into exactly four jets using one of the existing
jet finder algorithms.
A priori there are three possible ways to group the four jets into
two pairs of jets. The final pairing choice is achieved by a fit
procedure where the pairing with the highest fit probability is
taken to be the correct one.  
\begin{table}[htbp]
 \begin{center}
 \begin{tabular}{|ll|c|c|c|} \hline
      &  Final state & BR (\%)    & Purity(\%) & Eff.(\%) \\
\hline
      &                &                  &                 &                 \\ 
a)    &  $\WWqqqq$     &       46         &      80         &       85        \\
%     &                &                  &                 &                 \\ 
\hline
      &                &                  &                 &                 \\ 
b)    &  $\WWqqln$     &        44        &      90         &       80        \\
%              &                  &                 &                 \\  
\hline
      &                &                  &                  &                   \\ 
c)    &  $\WWlnln$     &       10         &      90          &       70        \\
%      &                &                  &                 &                  \\ 
 \hline                                                
  \end{tabular}
 \end{center}
\caption{The WW decay characteristics at the LEP2 experiments.}
\label{tb_ww}
\end{table}
At LEP2 energies the two W-bosons decay within a distance of about $0.1$ fm,
smaller than the hadronisation scale of $\sim$ 1.0 fm.
Thus the fragmentation process of
the two hadronic decay products from the two W-bosons in the 
same event are not necessarily independent.
Two main sources of possible interconnection between 
the two W-bosons decay products have been considered, Bose-Einstein 
correlations and the colour reconnection.
  
\subsection{Bose-Einstein correlation in WW decays}

\vspace*{2mm}

\begin{figure}
\begin{minipage}[0.0]{6.7cm}
\begin{center}\mbox{\input epsf \epsfysize 8.0cm
                        \epsfbox{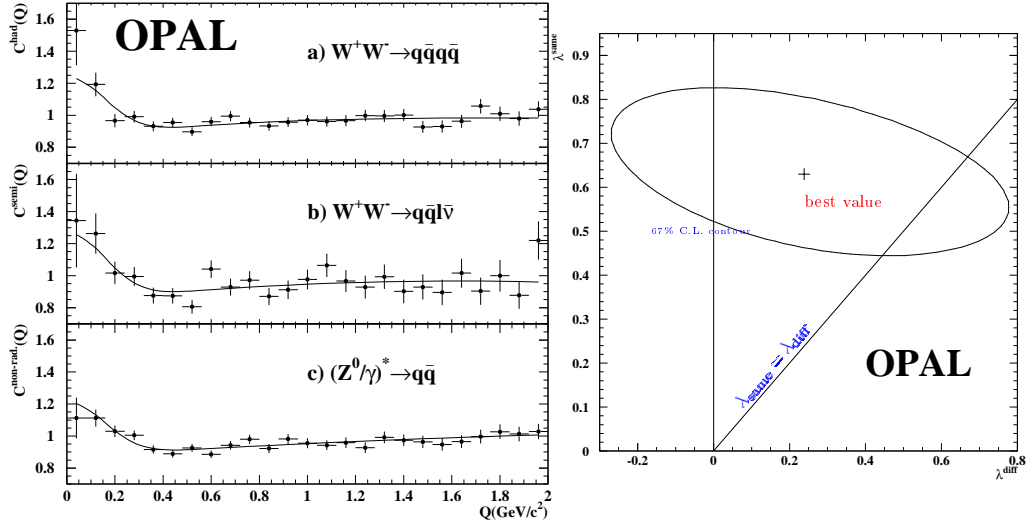}}\end{center}
\end{minipage}
\begin{minipage}[0.1]{6.5cm}

\vspace*{2.2cm}

\begin{center}\mbox{\input epsf \epsfysize 8.0cm
                        \epsfbox{procga9.eps}}\end{center}
\end{minipage}

\vspace*{-0.1cm} 
 
\caption{[left]The Correlation function in \eeWW 
events for like-charge pion pairs 
from the fully
hadronic, semileptonic and non-radiative events samples.
[right] The fit correlation between $\lambda_{same}$ and
$\lambda_{dif\!f}$ obtained by OPAL for the channel e$^+$e$^- \to\WWqqqq$.}
\label{fig_lsameldiff}
\end{figure}

\noindent
In dealing with a possible two-pion BEC effect on the M$_W$ measurement, 
it is useful first to investigate the BEC separately in the three 
decay channels defined in Table \ref{tb_ww} where they are listed
together with their branching ratio (BR) values and the purity and
efficiency achieved for them in the LEP experiments.\\ 
Typical $C_2(Q) = N^{++}(Q)/ M_c^{++}(Q)$
distributions for these three channels, obtained by OPAL \cite{opal_wwbec}
from the LEP2
data, are shown in Fig. \ref{fig_lsameldiff}[left] where low $Q$
enhancements is seen in all the channels. In Table \ref{tab:ww_bec1}
we show the BEC parameters obtained
by the ALEPH \cite{aleph_wwbec}, DELPHI \cite{delphi_wwbec} 
and OPAL \cite{opal_wwbec} collaborations  
from fits of an equation of the type: 

\vspace{-1mm}
 
\begin{center}
\begin{equation}
\label{becww_fit}
C_2(Q) \ = \ \kappa (1 + \lambda_2e^{-Q^2r^2_2})(1 + \varepsilon Q + \delta Q^2)
\end{equation}
\end{center}
to their data. Here ~$r_2$ ~is the radius of the pions source and $\lambda_2$,
which varies between 0 and 1, measures the strength of the two-pion BEC effect.
The term $(1 + \varepsilon Q + \delta Q^2)$ accounts for long range
correlations and $\kappa$ is a normalization factor.

\begin{table}[htbp]
 \begin{center}
 \begin{tabular}{|ll|c|c|c|} \hline
              &        & {\bf ALEPH }   &{\bf DELPHI} & {\bf OPAL} \\
\hline
              &         &              &              &  \\ 
$\WWqqqq$ & $r_2$(fm)     & $0.57 \pm 0.10 \pm 0.03$ & $0.57 \pm 0.12  \pm 0.06$ & $0.91 \pm 0.11 \pm 0.10$  \\
          & $\lambda_2$ & $0.18 \pm 0.04 \pm 0.04$ & $0.24 \pm 0.08 \pm 0.04$ & $0.43 \pm 0.15 \pm 0.09$   \\    
\hline
              &         &              &              &  \\ 
$\WWqqln$ & $r_2$(fm)     & $0.74 \pm 0.25 \pm 0.03$ & $ 0.49 \pm 0.07 \pm 0.02 $ & $0.91 \pm 0.11 \pm 0.10$ \\
          & $\lambda_2$ & $0.23 \pm 0.10 \pm 0.04$ & $ 0.43 \pm 0.09 \pm 0.04 $& $0.75 \pm 0.26 \pm 0.18 $  \\  

\hline
              &         &              &              &  \\ 
$(\Zzero/\gamma)^\ast \to q \bar q$ &$r_2$(fm)      &$--$ & 0.65 (fixed)   & $0.91 \pm 0.11 \pm 0.10$  \\
                       & $\lambda_2$ & $--$ &  $0.31 \pm 0.04 $  & $0.49 \pm 0.11 \pm 0.08$                  \\  
 \hline                                                
  \end{tabular}
 \end{center}
\caption{The BEC parameters determined in the 
\eeWW and $e^+ e^- \to (Z^0/\gamma)^{\ast} \to q \bar q$ samples by the
ALEPH, DELPHI and OPAL experiments at LEP2.}
\label{tab:ww_bec1}
\end{table}
\noindent
As can be seen from Table \ref{tab:ww_bec1}, significant two-pion BEC
exists in the fully hadronic events. In the semileptonic events, due to
the relative low statistics, the $\lambda_2$ values are only about 3 s.d.
away from zero. To note is that DELPHI and OPAL have also examined the
BEC in the non radiative $e^+ e^- \to (Z^0/\gamma)^{\ast} \to q \bar q$ 
events. In both experiments a $\lambda_2$ significant from zero was
found. Whereas DELPHI chose for the fit a  
fixed ~$r_2$ value of 0.65 fm, OPAL obtained
from their fit an ~$r_2$ ~value consistent with those obtained with data at
$e^+ e^-$ energies on the Z$^0$
mass and below.\\
Important to note is that the BEC results for the semileptonic WW data
are coming from the same W whereas the crucial question for the M$_W$ 
measurement is whether BEC exist also for pions coming from different
Ws. A handle on this question can be obtained from a comparison of
the BEC effects in the fully hadronic sample and in the semileptonic
sample. To this end one can define a probability density $P^{(4q)}(Q)$,
as a function of $Q$, for the two identical pions in the fully hadronic ($4q$)
WW events. In a linear scenario one has
$$P^{(4q)}(Q) = P^{(4q)}_s(Q) + P^{(4q)}_d(Q)$$
where  $P^{(4q)}_s(Q)$ and  $P^{(4q)}_d(Q)$ are respectively the
probability
density of two identical pions from the same W and from 
different Ws in the event. Using a MC generated sample without BEC
effects and assuming independent decays of the Ws
one can form a correlation function for pion-pairs coming from
different Ws  
$$R^{(4q)}_d(Q) = \frac{P^{(4q)}_d(Q)}{P^{(4q)}_{0d}(Q)} =
\frac{P^{(4q)}(Q) - 2 \cdot P^{(2q\ell \nu)}(Q)}{P^{(4q)}_{0d}(Q)}$$
This correlation function will be zero if there are no BEC between
pion-pair
of different Ws (see e.g. \cite{delphi_wwbec}).
Assuming the form $P_i^{(4q)}(Q) = P_{0i}^{(4q)}(1 + \lambda_i e^{-r_i^2Q^2})$
for $i= s$ or $d$ another correlation function can also be studied,
namely 

$$R^{(4q)}(Q) = \frac{P^{(4q)}_s(Q) + P^{(4q)}_d(Q)}
{P^{(4q)}_{0s}(Q) + P^{(4q)}_{0d}(Q)} = 
1+\left [\lambda_s - g(Q)(\lambda_s - \lambda_d) \right]e^{-r^2_2Q^2}$$
%in the last step $r_s$ = $r_d$ is assumed.
where $g(Q)$ is equal to: \ $g(Q) = P^{(4q)}_{0d}(Q)/
[P^{(4q)}_{0s}(Q) + P^{(4q)}_{0d}(Q)]$. In the last step $r_2 \equiv r_s$ = $r_d$
is assumed. This basic correlation function can further be expanded
and modified to
include the WW semileptonic data and the contribution from
the non-radiative $e^+ e^- \to (Z^0/\gamma)^{\ast}
\to q \bar q$ data which infiltrated in the selection of the WW events. 
%----------------------------------
\begin{table}[htbp]
 \begin{center}
 \begin{tabular}{|ll|c|c|c|} \hline
    &         & same W          & diff W              & $(\Zzero/\gamma)^\ast$ \\
\hline \hline 
 L3 & $r_2$(fm)   & $0.79 \pm 0.27$ & $ 1.03 \pm 0.85 $   & --       \\                                  
   &$\lambda_2$ & $0.32 \pm 0.17$ & $ 0.75 \pm 1.80 $   & --       \\
\hline
 OPAL &  $r_2$(fm)& $0.92 \pm 0.09 \pm 0.09$ & $0.92 \pm 0.09 \pm 0.09$ & $0.92 \pm 0.09 \pm 0.09$   \\                                     
   &$\lambda_2$ & $0.63 \pm 0.19 \pm 0.14$ & $0.22 \pm 0.53 \pm 0.14$   & $0.47 \pm 0.11 \pm 0.08$   \\
 \hline                                                
  \end{tabular}
 \end{center}
\caption{BEC parameters results from the fit to the WW events.}
\label{tab:ww_bec2}
\end{table}
The four LEP experiments have, so far, adopted somewhat different
approaches, which are related to the
correlation function described above,
to extract information on the $\lambda_d$. 
DELPHI and ALEPH, in their preliminary study 
compared their results for the BEC in the fully hadronic WW data
at 183 GeV (DELPHI) and 172+183 GeV (ALEPH) , 
with
MC generated samples having two-pion BEC in the same and different W
bosons. Their studies have indicated that $\lambda_d$ is consistent
with zero. 
The L3 collaboration \cite{l3_wwbec} 
used a fitting procedure to obtain the BEC parameters
$\lambda_s$, $r_s$ 
and $\lambda_d$, $r_d$
which are presented in Table \ref{tab:ww_bec2}. In the analysis of
OPAL an expanded version 
of the correlation function described above has
been used where also non-pion contamination has been considered. 
However due to the relative small statistics, $r_s = r_d =
r_{(Z^0/\gamma)^{\ast}}$ had to be assumed.
The OPAL fit results are also presented in Table \ref{tab:ww_bec2}. As
seen from this table, the values obtained by L3 and OPAL for $\lambda_d$ are non-zero
but due to the small statistics there are associated with large errors
and therefore are also consistent with zero. This situation is best
illustrated in Fig  6 [right] where the 67\% confidence level
contour for the OPAL $\lambda_{same}$ versus $\lambda_{dif\!f}$ fit
results are shown together with the best value. From this figure one
concludes that with the present statistics even values of
$\lambda_{dif\!f} > \lambda_{same}$ cannot be excluded. 
  
\subsection{Colour reconnection in WW decays}

\vspace*{2mm}

\noindent
%The products of
%the two W decays may have a significant space-time overlap since the
%separation of their decay vertices at LEP2 energies is small compared
%to characteristic hadronic distance scales. In the fully hadronic
%channel this may lead to final state interactions (FSI). 
In addition to the two-pion BEC discussed above, another
type of FSI  may affect the data namely 
the colour reconnection phenomenon.  This 
leads to a rearrangement of the colour flow between the two W bosons.
A variety of models, implementing interconnection effects, have been proposed
\cite{modelcr,ariad} which predict different biases in the measurement of the 
M$_W$ from the $\WWqqqq$ channel.
The extraction of the effect from the  data is usually done by  
comparing various properties of
the 4q and qql$\nu$ final state samples such as 
the charged multiplicity and average scaled momentum distributions in
addition to
global variables such as the mean thrust and the mean rapidity.
All the four LEP experiments \cite{cr,nigel}  have recently looked for
colour reconnection  effects in their \eeWW data 
collected with a total luminosity of  $\sim 55$~pb$^{-1}$.  
%%%%%
Their results for the charge multiplicity comparison between
the fully hadronic and the semileptonic \eeWW decays
are summarised in Table \ref{tab:ww-nch}. The quantity  
$$\Dnch = \nchQQQQ - 2 \nchQQLV$$ which is given in the table
is expected, in the absence of CR, to be
equal to zero. 
\begin{table}[htbp]
 \begin{center}
 \begin{tabular}{|l|c|c|c|} \hline
 Experiment &  $\nchQQQQ$ & $\nchQQLV$ & $\Dnch$ \\
\hline
 ALEPH$^\star$ & 
 $35.33 \pm 0.73$ & $17.01 \pm 0.37 $ & $1.31 \pm 0.83$   \\
 DELPHI  & 
 $37.36 \pm 1.00$ & $ 19.48 \pm 0.73 $   & $-1.6 \pm 1.5 $   \\
 L3 &                                   
  $36.3 \pm 0.9$ & $ 18.6 \pm 0.6 $   & $-1.0 \pm 0.9 $   \\
 OPAL &                                      
 $39.4 \pm 1.0$ & $ 19.3 \pm 0.4 $   & $0.7 \pm 1.0 $   \\
 \hline                                                
  \end{tabular}
  \end{center}
%\begin{center}
\vspace{-2mm}

\hspace{2.1cm}  $\star$ only partially unfolded.
%\end{center}
\caption{Mean charged particle multiplicities of the WW events
obtained by the LEP experiments.}
\label{tab:ww-nch}
\end{table}
All the errors listed in the table are the statistical and systematic 
ones added in quadrature.
Since the ALEPH mean multiplicity values are only
partially unfolded, they should not be directly compared
with the other three experimental results. Therefore it is more
appropriate to compare the $\Dnch$ values obtained by the four 
LEP experiments. 
The LEP average multiplicity difference  obtained,
using the $\Dnch$ values from the last table column, is: 
$${\Dnch}_{average}^{LEP} =  {\nchQQQQ} - 2{\nchQQLV} = +0.20 \pm 0.50
\ (stat.+syst.)$$
This result is consistent with the absence of any change in the $\nchQQQQ$ due to colour
reconnection.
%%%%
Essentially all the $\SK$, HERWIG and ARIADNE models \cite{modelcr,ariad} are consistent with
data. Some more extreme models however, such as the
instantaneous reconnection scenarios in the  $\SK$ model, and the
ARIADNE model AR 3~\cite{ariad}, in which gluons having
energies greater than $\Gw $ are allowed to interact, are disfavoured.\\
%%%%%%
DELPHI and OPAL \cite{cr,nigel} measured also the dispersions of the charged multiplicity
distributions, $\DQQQQ$  and $\DQQLV$, and their difference
given by $\DD=\DQQQQ-\sqrt{2}\DQQLV$ . The results are consistent with Monte Carlo
expectations. Furthermore
there is no indication for differences in charged multiplicity distribution's
shape based on the first moments of the multiplicity distributions. \\
Other studies like fragmentation functions and trust distributions show no significant
difference between the data in the two channels of the W-pair decays \cite{nigel}.
Thus we may conclude that at the present level of statistics no evidence for colour
interconnection effects has been observed in $\WWqqqq$  events.
 
\subsection{M$_W$ uncertainties from FSI}

\vspace*{2mm}

\noindent
In the absence of any experimental evidence for the existence of FSI  between the
decay  products of different W-bosons, the systematic uncertainty from such 
effects is for the time being estimated from Monte Carlo models \cite{Wmass}.
The DELPHI, L3 and OPAL collaborations quote for M$_W$
measured in the channel $\WWqqqq$ a combined systematic uncertainty
 of 100 MeV for BEC and CR. 
The ALEPH Collaboration quotes 25 MeV for CR and 50 MeV for  BEC. \\
Thus the combined LEP result for the W-boson mass obtained  
from direct reconstruction and threshold measurements using both the
fully
hadronic and semileptonic data is
currently evaluated to be  \cite{Wmass}: \\
$$\rm{M}_W \  = \ 80.37\ \pm \ 0.07(stat.+syst.)\ \pm \ 0.04(FSI)\ \pm \
0.02(E_{beam}) \ 
\ \rm{GeV}$$  
The reduction from $\sim$ 90 GeV to 40 GeV of the uncertainty coming
from FSI is due to the fact that the final M$_W$ is coming also from
the semileptonic sample where no FSI are present.   
%\newpage
\section{The $\Lambda \Lambda$ emitter dimension}

\vspace*{4mm}

\noindent
The two-pion emitter dimension, ~$ r_2(\pi\pi)$, has been measured via the
BEC in a
large variety of interactions and over a large range of energies. At the
same time essentially no information has been  
available on the emitter
dimension of two identical baryons in multihadronic final states. 
Recently it has been pointed out
\cite{lipkin} that this di-baryon emitter radius can be evaluated through
the onset of the Pauli exclusion principle as the difference of
the baryons four momenta, $Q$,
decreases to zero. 
To this end a method for the determination of 
the spin composition of the
$\Lambda\Lambda$$(\bar{\Lambda}\bar{\Lambda})$ pairs has been 
developed in \cite{lipkin}, 
which is based on the data themselves without the need of a reference
sample.\\

\noindent
In this method each $\Lambda$ of the hyperon pair 
is first transformed to their common CM system and then each decay proton
(anti-proton) is transformed to its parent hyperon (anti-hyperon) CM system.
The distribution of the cosine angle between these two protons, here denoted by
$y^{\ast}$, depends on the fraction of the S=1 (or S=0) in the data.
Specifically the following $y^{\ast}$ distributions are expected 
for pure S=0 and S=1 states of the $\Lambda$ pair 
%\cite{alex_lipkin}
\vspace{-2mm}
%\begin{flushleft}
$$dN/dy^{\ast}|_{_{S = 0}} \ = 1 + (-1)^{B/2} 
\cdot \alpha_{\Lambda}^2 \cdot y^{\ast} \ \ \ \ \rm{and} \ \ \ \
%\hspace{2cm}
dN/dy^{\ast}|_{_{S = 1}} \ = 1 - (-1)^{B/2}
\cdot \alpha_{\Lambda}^2 \cdot y^{\ast}/3$$

\vspace{1mm}

\noindent
where $\alpha_{\Lambda}$=0.642$\pm$0.013 \cite{pdg98}
is the $\Lambda \to p \pi^-$ decay parameter arising from parity violation
and B is the baryon number of the di-hyperon system.
These distributions are independent of the orbital angular 
momentum and are valid as long the $\Lambda$'s are non-relativistic
in their di-$\Lambda$ CM system. Therefore the spin composition 
analyses in this method are
restricted to relatively small $Q$ values\footnote{Assuming that no 
$\Lambda \Lambda$ resonances exist in this Q region.}. 
Defining $\varepsilon$ as the fraction of S=1 
in the di-hyperon sample, namely
 $$\varepsilon =\frac{(S=1)}{(S=1)+(S=0)}$$ 
then the following function can be fitted to
the data:
\vspace{-1.5mm}

%\begin{flushleft}
$$dN /dy^{\ast}=f_{BG} \ + \
(1-f_{BG})\cdot {\{}(1-{\bf \varepsilon})\cdot dN / dy^{\ast}|_{_{S = 0}}
+\varepsilon \cdot dN / dy^{\ast}|_{_{S=1}}{\}}$$
%\end{flushleft}
\vspace{-2mm}

\noindent
where $f_{BG}$ is the background fraction in the data. 
%This background was found
%to be approximately independent of $y^{\ast}$ and equal to 
%about 10\% and 60\% in the $\Lambda \bar \Lambda$ and $\Lambda 

\noindent
The fraction of the S=1 spin content, $\varepsilon$,  as a function of $Q$ has been measured 
in $\sim 4 \times 10^6$  hadronic $\Zzero$ decays per experiment at LEP  
by the OPAL \cite{Lam1}, ALEPH \cite{Lam3} 
and DELPHI \cite{Lam2} collaborations.
The R$_{\Lambda \Lambda}$ values obtained by 
the OPAL and DELPHI collaborations are : \\
\[ {\bf OPAL(96)} ~~~~\large  R_{\Lambda\Lambda}
 = 0.19 ^{+ 0.37}_{-0.07} \pm 0.02 ~~{\mathrm fm} \] \\

\vspace*{-5mm}

\[ {\bf DELPHI(98)} ~~ \large R_{\Lambda\Lambda}
 = 0.11 ^{+ 0.05}_{-0.03} \pm 0.01 ~~{\mathrm fm} \] 
The ALEPH results for the $\Lambda\Lambda$ spin composition
 are similar to those of OPAL and DELPHI 
but no attempt has been made to extract from them an 
$R_{\Lambda \Lambda}$ value. 
 
%\vspace{-4mm}

\begin{figure}[htbp]
\begin{minipage}[0.0]{6.5cm}
\begin{center}\mbox{\input epsf \epsfysize 7.3cm
                        \epsfbox{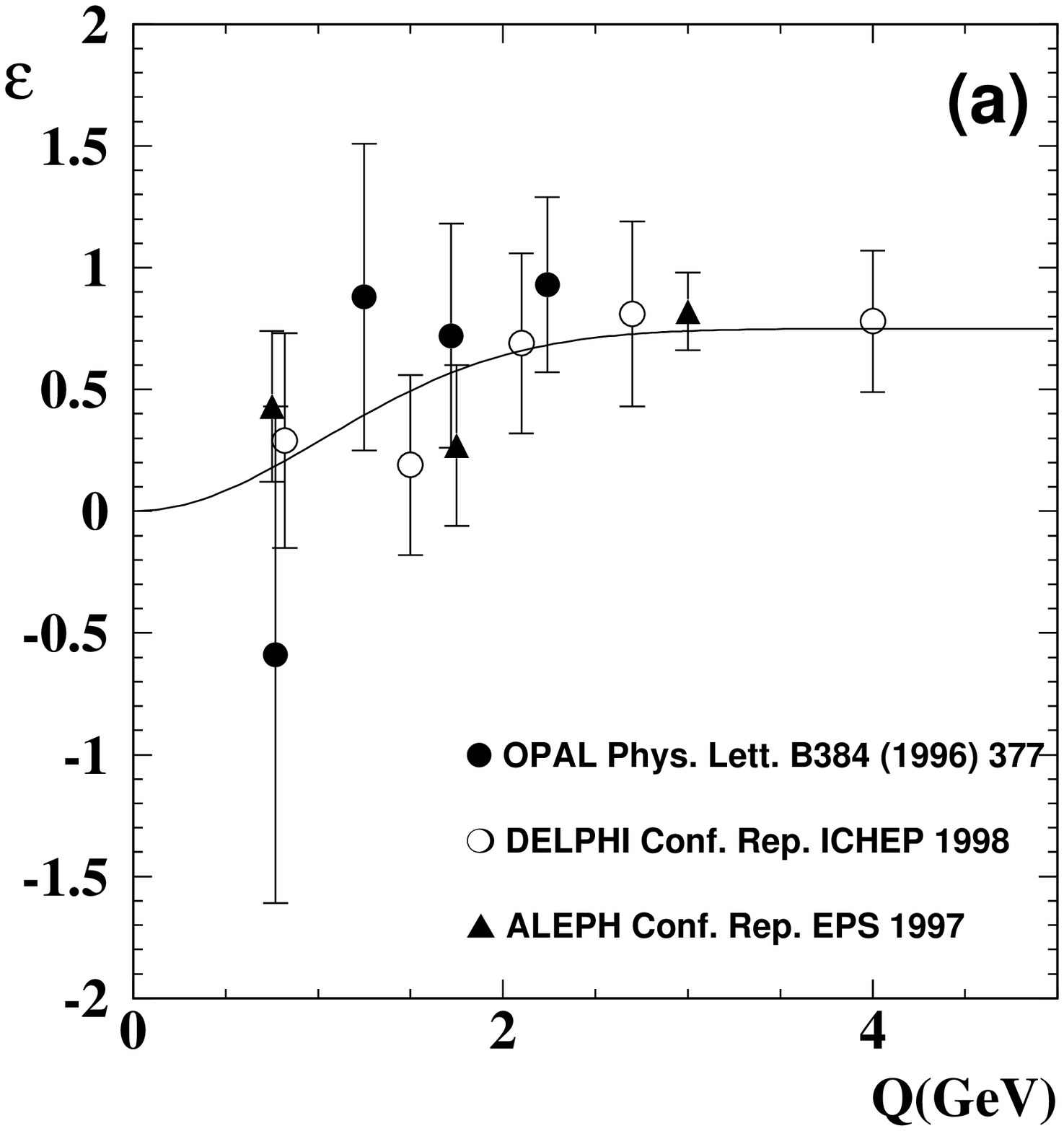}}\end{center}
\end{minipage}
\begin{minipage}[0.0]{6.5cm}
\begin{center}\mbox{\input epsf \epsfysize 7.1cm
                        \epsfbox{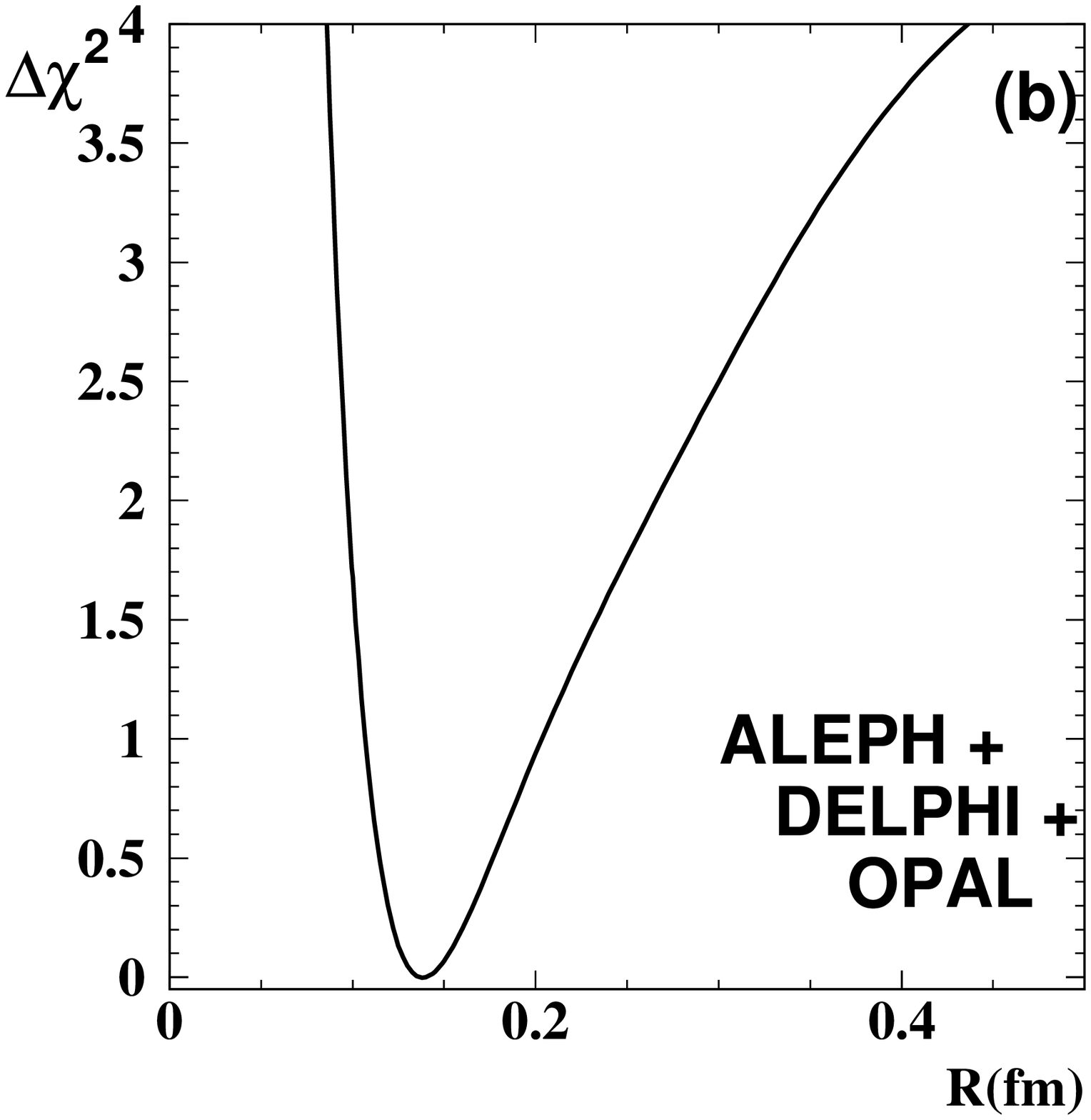}}\end{center}
\end{minipage}

\vspace*{-2mm}

\caption{[a]The S = 1 fraction, $\varepsilon$, of the 
$\Lambda\Lambda(\bar \Lambda \bar \Lambda)$ pairs 
measured as a function of $Q$ by the ALEPH, DELPHI and OPAL 
collaborations. The solid line represents the results of the fit of
Eq. \ref{pauli} to the data points. 
[b] The  $\Delta\chi^2 = \chi^2 - \chi^2_{min}$ dependence on
$R_{\Lambda\Lambda}$ . }
\label{fig_lamlam}
\end{figure}  

\noindent
The measured  $\varepsilon(Q)$ values of the three experiments
are compiled in Fig. \ref{fig_lamlam}a.     
The solid line in the same figure  is the outcome of an
unbinned maximum likelihood fit of the function:
% in the range $0.5 < Q < 2.5 GeV to the function:

\vspace*{-4mm} 

\begin{center}
\begin{equation}
\varepsilon(Q) = 0.75 
    [1- e^{(-R^2_{\Lambda\Lambda}Q^2)}]
\label{pauli}
\end{equation}
\end{center}
to the data points plotted
in the figure. The constant
0.75 appearing in the function represents a statistical spin mixture
which is proportional to 2S + 1.  
From this over-all   
fit the dimension of the $\Lambda \Lambda$ emitter and its
errors are found to be 
\begin{center}  
\[ R_{\Lambda\Lambda} = 0.14 ^{+ 0.07}_{-0.03} ~~{\mathrm fm} \]
\end{center}
%\noindent  
The $\Delta\chi^2 = \chi^2 -\chi^2_{min}$ behaviour of this fit is shown in 
Fig. \ref{fig_lamlam}b.
The ALEPH collaboration has also used  
an alternative method to measure  $R_{\Lambda\Lambda}$. In that method
one constructs, similar to the BEC studies, a correlation function of
the type
$$C_2^{\Lambda \Lambda}(Q) = N^{\Lambda \Lambda}_{data}(Q)
/  N^{\Lambda \Lambda}_{ref}(Q)$$  
where the numerator is the data distribution and the denominator
is the distribution of a reference sample void of the Fermi-Dirac
statistics. This $C_2^{\Lambda \Lambda}(Q)$ correlation is expected to decrease
at low $Q$ values due to the onset of the Pauli exclusion principle.
From this analysis ALEPH evaluates  $R_{\Lambda\Lambda}$ to be
$(0.10 - 0.09) \pm 0.02$ fm in perfect agreement with the values
obtained from the spin composition analyses.\\   

\noindent
Comparing $R_{\Lambda\Lambda}$ with former measured $r_2(KK)$ \cite{KKdelphi}
and $r_2(\pi \pi)$ \cite{wolf} 
values it is interesting to note that the larger the mass of the
emitted particle the smaller is the dimension of its source. 
\section{Conclusion}

\vspace*{4mm}

\noindent
The LEP studies of BEC have been extended by
OPAL to one-dimensional genuine identical charged three-pion system.
These genuine BEC reveal a significant enhancement near threshold 
which may warrant their implementation in the MC models. Furthermore
it may be thus important to extend these studies to higher
orders. The L3 collaboration has extended the two-pion BEC studies to
two dimensions. Their results indicate that the emitter shape is not
spherical. Extension of this study to three dimensions is straight forward
and of  considerable interest.
So far no experimental conclusion can be derived for the contribution,
if any, of the BEC and/or colour reconnection on the $\WWqqqq$
channel. Thus the systematic 
error due to FSI associated with the measured M$_W$ are estimated from model
dependent MC studies. It is expected however that by the year 2000
sufficient data will be
available to extract from them the FSI effects
on the measured W mass. Recently information is also gathered on the
baryon emitter dimension. This should be further explored, not only
experimentally but also theoretically.

\section*{Acknowledgments}

\vspace{3mm}

\noindent
%We particularly would like to thank the SL Division CERN for the
% efficient operation of the LEP accelerator at all energies and
%for their continuing close cooperation with the experimental groups. 
We would like to thank the members of the four LEP experiments 
for providing the information needed to prepare this report.

\end{document}